\documentclass[12pt]{iopart}
\usepackage{graphicx}
\usepackage{hyperref}
\begin{document}

This is the version of the article before peer review or editing, as submitted by an author to the European Journal of Physics. IOP Publishing Ltd is not responsible for any errors or omissions in this version of the manuscript or any version derived from it. The Version of Record is available online at \url{https://doi.org/10.1088/1361-6404/ac7830}.

\title{Collaborative smartphone experiments for large audiences with phyphox}

\author{S Staacks$^1$, D Dorsel$^1$, S Hütz$^1$, F Stallmach$^2$, T Splith$^2$, H Heinke$^1$, C Stampfer$^1$}

\address{$^1$ Institute of Physics I and II, RWTH Aachen University, Aachen, Germany}
\address{$^2$ Didactic of Physics, Leipzig University, Leipzig, Germany}
\ead{staacks@physik.rwth-aachen.de}

\begin{abstract}
We present methods to implement collaborative experimentation with smartphone sensors for larger audiences as typically found at Universities. These methods are based on the app ``phyphox'', which is being developed by the authors, and encompass simple data collection via web forms as well as a new network interface for ``phyphox'', allowing to collect real-time experiment data from an audience on-site or easy data submission for remote participants. Examples are given with practical considerations derived from first implementations of this method in a lecture hall with 350 undergraduate students as well as a global experiment to determine the Earth's axial tilt with smartphones.
\end{abstract}

%
% Uncomment for keywords
%\vspace{2pc}
%\noindent{\it Keywords}: XXXXXX, YYYYYYYY, ZZZZZZZZZ
%
% Uncomment for Submitted to journal title message
%\submitto{\JPA}
%
% Uncomment if a separate title page is required
%\maketitle
% 
% For two-column output uncomment the next line and choose [10pt] rather than [12pt] in the \documentclass declaration
%\ioptwocol
%

\section*{Introduction}

Ever since smartphones have become ubiquitous among students, the sensors in these devices have been used for experimentation in science education \cite{vogt, pendrill, kuhn, chevrier, vieyra}. Allowing students to discover the world with their own measuring devices is not only considered to be a refreshingly unusual variation of student experimentation, but it is also a free and readily available chance to do quantitative measurements with digitally acquired data and digital data analysis. This aspect becomes even more relevant if dedicated measuring equipment is not available due to limited resources or because of disproportionate logistical requirements.

The latter in particular applies to larger courses in higher education. In the context of a lecture or its accompanying exercise courses, this organized experimentation quickly becomes impractical and for this reason student experimentation is rarely realized in the context of a lecture so far. In contrast, experimental assignments using the student‘s own devices are an engaging alternative to purely mathematical assignments \cite{stampfer, Kaps_2021}.

Surprisingly, while smartphones are mostly known and widely used because of their connectivity and networking capabilities, smartphone experimentation so far rarely takes advantage of this. Aside from few dedicated apps that aim at data collection for citizen science projects \cite{odenwald, lemmens}, typical smartphone experiments only use the data collected on a single device or require a manual export and subsequent merging of the data in a separate analysis tool. In contrast, the phone’s connectivity could be used for more accessible and engaging collaborative experimentation across a large audience.

In this paper we report on a new network interface for the education-centric sensor app ``phyphox'' \cite{staacks} with examples for different learning situations and scenarios in which this new interface has been used. These include real-time collaborative experiments in a lecture hall with hundreds of undergraduate university students and a real-time collaborative experiment with users around the globe in an informal learning setting.

The paper is structured to incorporate additional challenges or peculiarities with each example, starting with (a) an example for manual data collection not yet using the new interface to demonstrate how experimentation in a lecture context can be used to flip the experiment experience for physics students. Then (b) the new interface is introduced, followed by (c) an example on how it is used in the same lecture to transform the concept into a real-time experience for students. In examples (a) to (c) the physics students were asked to use their smartphones as an ocillating balance to determine the unknown mass of different objects \cite{Kaps_2020}. We then discuss with example (d) the requirement to filter data from incorrectly conducted experiments to go on to (e) an example of data collected from even less reliable experimentators as users across the globe collect data to determine Earth‘s axial tilt. Finally, we present (f) the requirements to implement the network interface in other courses.

\section*{a. Flipping the class room with a web form}

In order to perform a collaborative experiment with hundreds of students without a function to directly submit experiment data from within the experiment app, a simple web form can be used. In contrast to other methods like collecting results via email, a dedicated form forces students to enter their data in a given machine processable format, allowing for easy scaling to hundreds of students without an increased effort to manually merge differently formatted emails.

We employed this solution in the lecture ``experimental physics 1'' of the winter term 2019/2020 for about 350 first semester students aiming for a Bachelor’s degree in physics at the RWTH Aachen University. The assignment was part of the compulsory exercises that accompany the lecture, but this particular assignment was considered optional, allowing the students to earn bonus points which would allow them to leave out a different assignment.

The idea was to flip the typical experiment experience known from most lectures on experimental physics. Instead of letting the lecturer derive a theory and confirm it with an experiment on stage, we aimed to let the students conduct the experiment without prior knowledge of its expected outcome and let the lecturer explain the data in a later lecture.

A good experiment for this is a simple gravity pendulum as its construction only requires household items available to most of the students. The students were instructed to build a pendulum using their smartphones, a piece of string and a small plastic bag or paper roll to hold the phone. We specifically asked them to use a swing-like suspension with two or four strings as seen in figure 1a instead of a single string to avoid an additional rotation about the axis of the single string. This way, the pendulum experiment configuration included in phyphox can be used to precisely determine the pendulum frequency.

\begin{figure}[h]
  \includegraphics[width=\textwidth]{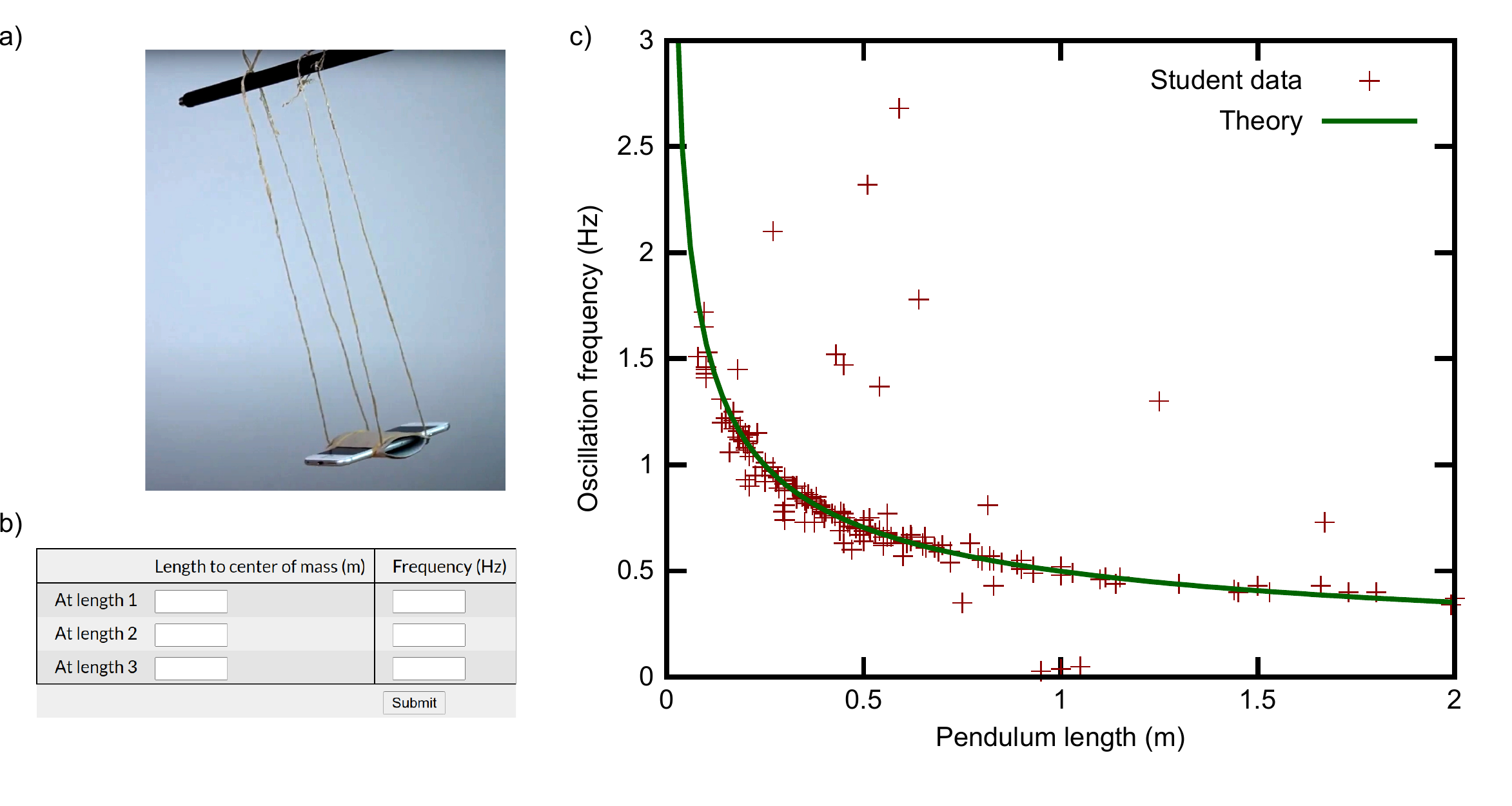}
  \caption{a) Example setup for a home-built smartphone pendulum. b) Screenshot of the translated submission form for the students. The original form was in German. c) Plot of pendulum frequency as a function of distance from pendulum axis to center of mass. A total of 195 data points submitted by 65 student groups from multiple years are plotted as well as the expected behavior of a mathematical pendulum. Few data points corresponding to larger constructions of up to $5\,\mathrm{m}$ (done in a stairwell) were left out in favor of scaling the axis such that the other data points can be well distinguished.}
\end{figure}

The assignment was done by students in groups of two or three and each group should repeat the experiment for three different lengths of string. The resulting data pairs of length and frequency were collected with a web form (figure 1b) that was integrated in the lecture’s digital script which in turn was embedded in the lecture’s virtual room on the University’s learning management system.

The experimental results were due before the lecture reached the topic of oscillators, so the lecturer, was able to use the student’s data in this particular lecture. After deriving and solving the equation of motion for a mathematical pendulum, he could compare the theoretical expectation with the collective data of the students, mimicking the scientific method and using the student’s unbiased experimental data. This procedure replaced the part of the lecture when typically an experiment with very few data points was conducted on stage.

We received data sets from 49 student groups, who each submitted three pairs of pendulum length \textit{l} and measured frequency \textit{f}. While most pendulums were limited to a string length of less than $2\,\mathrm{m}$, some setups in stairwells featured string lengths of up to $5\,\mathrm{m}$, exhibiting a strong motivation for the students. As shown in figure 1c, despite few outliers, the data set generally reproduces the theoretically expected behavior of $2\pi f = \sqrt{g/l}$ very well. 

\section*{b. Implementation of a network interface for automated data collection}

Although this approach had worked reliably and was well received by the students since we had first used it in winter term 2016/2017, there are some drawbacks to using a web form separate from the experimentation app. In particular collecting the data in one tool and then entering it in another one makes the procedure impractical for a real-time application. It also introduces an unnecessary hurdle, especially if used in informal learning settings where users are not familiar with an accompanying website like a digital lecture script.

Therefore we decided to implement a generic network interface to our data acquisition app ``phyphox'' \cite{staacks} to allow for an easy exchange of data by the students. Phyphox is add-free, open-source and specifically designed for science education. It is available for free on Android and iOS and features a fully documented file format that allows educators to design their own specific experiment configurations including choice of multiple smartphone sensors, experiment-specific data analysis, custom layout of graphs and visualizations, informative texts and user inputs like buttons or text boxes for numeric values. Every experiment configuration listed in the main menu of phyphox (figure 2a) is defined in this XML-based format. Educators can modify these or create entirely new experiment configurations specific for their classes or courses. These configurations can then be shared with students using QR codes (figure 2b). They can also be permanently integrated into the main menu of the locally installed phyphox app in a custom category and with a specific icon (figure 2c), thus allowing lecturers to build a section in phyphox that is specific to their course.

\begin{figure}[h]
  \includegraphics[width=\textwidth]{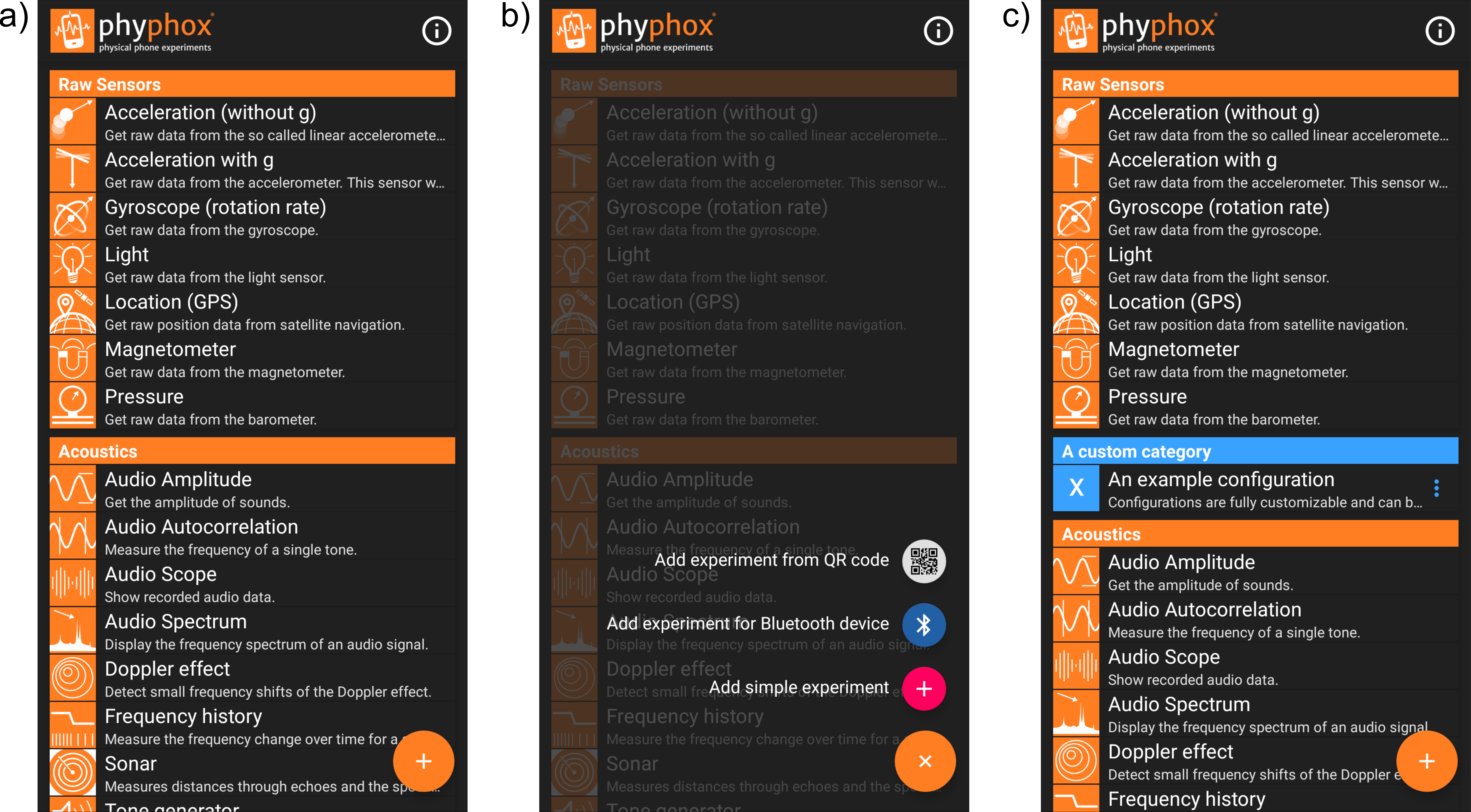}
  \caption{Series of screenshots of phyphox on Android. a) Main menu of phyphox after a clean installation. b) Menu to add new experiment configurations, in particular by scanning a QR code. c) After scanning a QR code, another configuration is available in the main menu. The blue color and the ``x'' as an icon are customizable as well as titles and labels.}
\end{figure}

The new network interface is also designed to be defined in this XML format. Since it is supposed to be open and compatible with as many server structures and protocols as possible, it allows to freely define which data is being sent and how data received from the network should be handled. Sensor data as well as user input or the result from the data analysis defined in the XML file can be sent. Conversely, data received from the network can be analyzed or displayed. Also, metadata like the make and model of the device, the phyphox version or a uniquely generated user id can be submitted.

To protect the users’ privacy, phyphox will inform the user about the data sources used in an experiment and that data will be submitted to a network service. Critical data sources like the microphone or GPS are indicated separately as well as critical metadata like the unique user id, which is unique only for a single service address to avoid tracking across services. A URL to a privacy policy can be added in the XML definition to inform the user about how the data is being processed.

At the moment, only a static server address can be configured and HTTP and MQTT are implemented as available protocols with variants like GET or POST methods for submission via HTTP and JSON or CSV payloads for MQTT. Phyphox and the XML format are designed such that new protocols can easily be implemented and non-static server addresses (for example discovered via mDNS) can be supported in the future.

For details about the exact configuration and current state of the network interface, please refer to the documentation on phyphox.org \cite{wiki} or its snapshot for the time of this article in the supplementary material.

\section*{c. Collaborative live experiments in a lecture hall}

The new network interface allowed us to do another oscillator experiment similar to the pendulum experiment in the same physics course but as a real-time experiment. In the context of being the last lecture before winter break we handed out a small clear plastic bag, a spring and chocolate bars to pairs of students. With these, students could construct a spring oscillator by putting the phone into the bag and attaching the bag to the spring (figure 3a). The chocolate bar could be used to modify the mass of the oscillator with additional $100\,\mathrm{g}$ per bar or fractions thereof that can easily be achieved thanks to the chocolate bar’s sections. The chocolate adds to the phone’s weight, which the students could measure with scales in the lecture hall or simply research online. Of course, the additional mass of the chocolate bars can be replaced by other low cost material like a set of appropriate metallic nuts or shims.

\begin{figure}[h]
  \includegraphics[width=\textwidth]{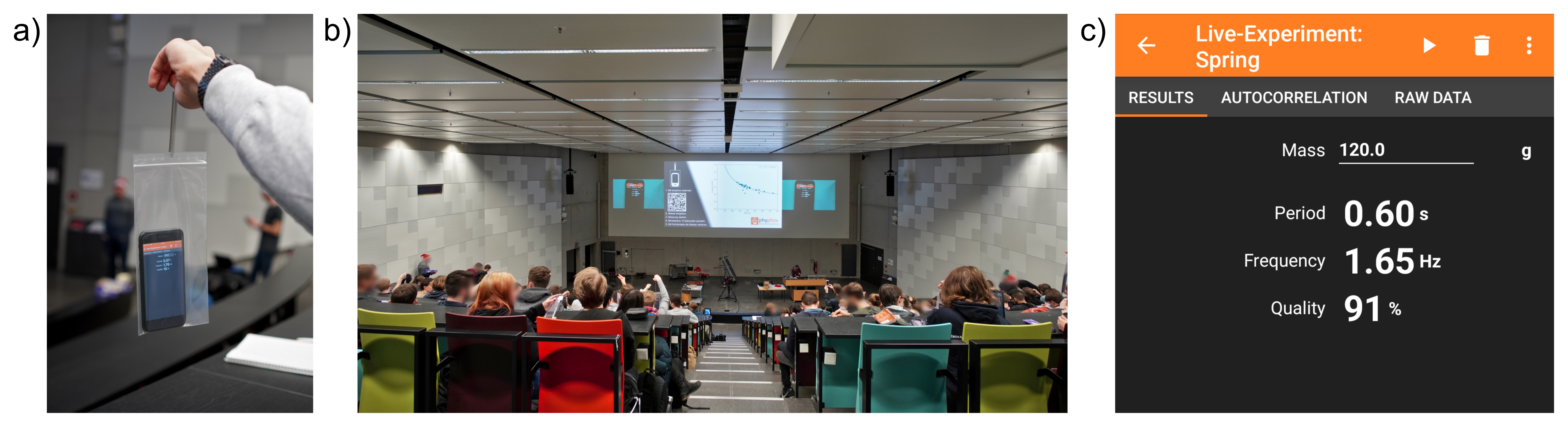}
  \caption{a) Example for a spring oscillator made from a spring and a plastic bag. b) A lecture hall participating in a live network experiment with collaborative results being presented on the main projector. c) Screenshot of the experiment configuration used in the collaborative spring experiment. Note that there is no submission button, but that results are being submitted periodically in this example.}
\end{figure}

The goal of the collaborative experiment was to generate a frequency over mass plot to determine the spring constant for the identical springs used by the students. In order to collect the data, a QR code was given to the students with a phyphox experiment configuration specific for this event. This configuration allowed the students to enter the current mass of their oscillator and recorded data from the accelerometer, which is continuously analyzed to determine the oscillation frequency from an autocorrelation (figure 3c).

Since the students could not easily press a button to submit data while the phone oscillates on the spring, the configuration was set up to periodically submit the current frequency to a server every ten seconds. A simple PHP script on the server was used to store the incoming data into a plain text file. To avoid duplicates and meaningless data points from handling the phone, the ratio of the maxima in the autocorrelation were used to get a measure for the quality of the data, which allowed the server to reject unusable data and to only keep a single data set per user id per oscillator mass.

Finally, the real-time experience of the experiment was achieved by a Python script running on the server, which periodically ran through the collected data to generate the frequency over mass plot. The result was displayed and updated every few seconds on the main projector of the lecture hall (figure 3b). Multiple variants of data representation were available, including a plot with a model fit to extract the spring constant, allowing the lecturer to switch between views and discuss the results as he moderated the ongoing experiment.

\section*{d. Filtering requirements for large data sets}

As the automated submission through the network interface along with the automated data analysis and presentation from the Python script allows to scale an experiment for large audiences while still being able to get real-time results, filtering incoming data becomes an important aspect. With an increasing number of participants the probability for erroneous data sets increases, which might interfere with the automated data analysis as fits are distorted by extreme outliers or the visualization might become unusable with erroneous data masking the actual measurement or throwing off the formatting as automated axes scale to extreme values. 

As larger audiences tend to be more anonymous, these erroneous data points can be intentional attempts by students to test the limits of the experiment, but in most cases simple mistakes can be sufficient to generate problematic data points. While we did not notice any intentional attempts to challenge our scripts, we noticed two common mistakes that highlight the importance to filter incoming data and the need for a suitable experimental design to avoid simple mistakes.

The first problem was anticipated and had been accounted for in the design of the experiment. As the phyphox configuration submits derived quantities of the sensor data every ten seconds without confirmation from the students, this will naturally include frequencies derived from noise while handling the phone and partially recorded oscillations if the analysis interval overlaps with starting or stopping the oscillator. These situations could reliably be detected by evaluating the ratio of the peak of the autocorrelation used to determine the frequency to the maximum of the autocorrelated data. For proper harmonic oscillations, this ratio will be close to 1.0 and any ratio below 0.75 is rejected. In the real-time analysis in the lecture for each user id only the result with the highest ratio was taken into account for each mass entered. Additionally, the accepted frequency range was $0.1\,\mathrm{Hz}$ to $4\,\mathrm{Hz}$ and the mass entered by the students was limited to a range of $50\,\mathrm{g}$ to $1\,\mathrm{kg}$.

While we did not see any noise from handling the phone passing this filter, we did not take into account simply forgetting to update the mass after varying it. The expected result of decreasing frequency with increasing mass can clearly be seen in the student data, but there is a significant number of additional seemingly chaotic data points in the bottom left corner influencing the fit. These data points occur if a student changes the mass of the oscillator without entering the new value into phyphox. The resulting measurement passes all filters, but the frequency will be associated with the wrong mass and even substitute the former correct measurement if a high ratio for the autocorrelation maxima is achieved. As the students start with only their phone as the mass of the pendulum and usually add additional mass later, this mistake will systematically move data points from the correct high frequencies with low masses to the bottom of the graph as the same low mass is now associated with the lower frequency of an oscillator that actually has a higher mass.

After repeating this oscillation balance experiment with a group of 50 first year physics teacher students during an experimental physics lecture in January 2019 at the University of Leipzig and observing the same mistake, we decided to group all data submissions for each student with the same mass and only take into account a single submission that was submitted in the middle of a series of identical mass values. The idea is that in most cases, the forgotten mass update would eventually be corrected by the students, so the wrong mass values would be at the end of the previous data series. Taking a value from the middle avoids early values that might include handling, but also cuts off contributions from a modified mass before its values has been entered.

\begin{figure}[h]
  \includegraphics[width=\textwidth]{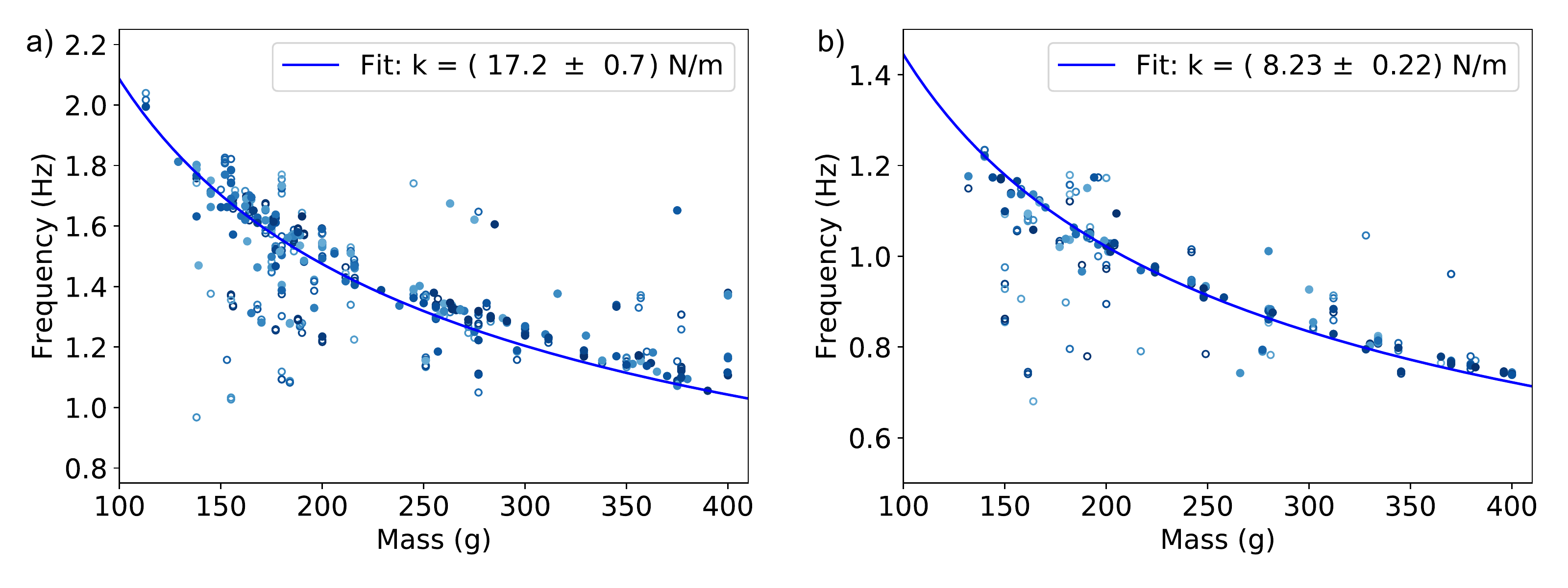}
  \caption{Result of the collaborative spring experiment as presented to the students during a lecture a) in Aachen and b) in Leipzig. Closed circles represent valid data points from students and open circles represent discarded data points. Shades of blue denote the confidence estimate from the ratio of maxima in the autocorrelation with light blue corresponding to 0.75 and dark blue corresponding to 1.0 (see text for details). Note, that different springs were used in both locations.}
\end{figure}

The other values have been discarded, but are still plotted as open circles in figure 4. The effectiveness of this strategy is supported by the observation that most outlying data points get eliminated this way. Note, that there are plenty of open circles at reasonable frequency/mass combinations that are discarded as well. These are additional valid measurements within a series submitted by the same user id and are therefore duplicates from the same individual experiment.

\section*{e. Global experiment to determine Earth's axial tilt}

Network-based collaborative smartphone experiments can be pushed to a global level and can produce astonishing results. If these experiments are performed with experimentators who are not part of a physics course, the requirements for proper filtering algorithms increase.

We conducted such an experiment on 21st December 2019, which was the day of the winter solstice, with the goal to trace the Sun’s path across the sky and thereby determine Earth’s axial tilt. To do this, we published a call to phyphox users through our website and social media channels to load a prepared experiment configuration (figure 5b) into phyphox. This experiment was designed to determine and submit the Sun’s position as seen from the user’s location multiple times throughout the day.

\begin{figure}[h]
  \includegraphics[width=\textwidth]{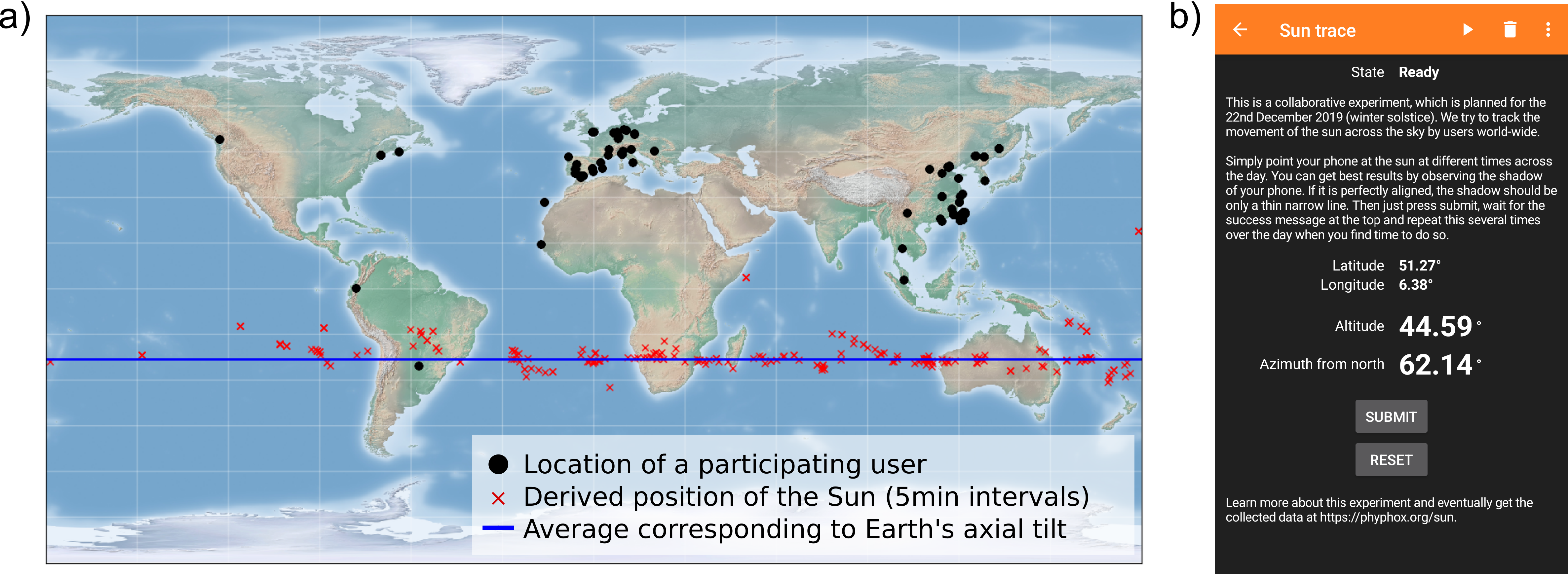}
  \caption{a) Global experiment to trace the Sun's location during the winter solstice. Black dots represent user locations, red crosses represent derived Sun locations and the blue line represents the average latitude corresponding to Earth's axial tilt. b) Screenshot of the experiment configuration used in the collaborative sun trace experiment. Note that this configuration features a submission button and only submits data if initiated by the user. (Location data has been shortened in the screenshot to protect privacy.)}
\end{figure}

To do so, the users have to align their phone with the Sun such that its \textit{y} axis (long side of the phone) is pointing directly at it. This can easily be achieved by observing the shadow of the phone and turning it slowly until the shadow becomes as small as possible, matching the cross-section of the phone. The experiment configuration that we provided then uses the magnetometer as a compass to determine the Sun’s azimuth and the accelerometer to determine its altitude by considering the orientation at which Earth’s acceleration occurs in the accelerometer’s frame of reference. The experiment configuration would also determine the user’s location which is submitted along with a unique user id and the determined azimuth and altitude when the user presses a button.

All submissions are collected on our server and a Python script was used to derive the Sun’s path from these submissions. However, the error rate we observed in the raw data is very high, which we attribute to users pressing the submit button out of curiosity without aligning their device with the sun, miscalibrated magnetometers or measurements near magnetic fields and misunderstandings of the experiment instructions like pressing the button after the phone has been moved away from its aligned position.

In order to filter these results in an unbiased way, our script takes into account data points that span the period of one hour. For each group of data points, the script then determines an average location for the Sun and removes the farthest outlier from that result. The remaining points are used again for a new average and removing another outlier, which is iteratively repeated until 50\% of the data points in the period have been removed. This method favors the data points in the middle of the one hour period and ideally the data points at the beginning and the end of the period will be removed, too. Such a long period still helps to stabilize the averaging until the worst outliers have been filtered while still yielding a much better temporal resolution than one hour. Therefore, this process has been applied repeatedly to one hour intervals that are shifted by five minutes each.

The result can be seen in figure 5a with several interesting features. As black points represent the location of a contributing user and red crosses the derived location of the Sun for each five minute shift, it becomes clear that we mostly have contributions from the northern hemisphere tracing the Sun that travels above a southern latitude. Unfortunately, there was some heavy overcast over Europe and parts of North America on that day, which made contributions from users in Central or Northern Europe very difficult and which might also have reduced the number of contributions from North America.

Still, aside from few deviations, the filtered Sun positions form a clear path and a fit (blue line) yielded an average latitude of $23.03^\circ \pm 0.07^\circ$, which matches the actual value of $23.4^\circ$ surprisingly well with only a little systematic skew to the North which we attribute to the fact that almost all measurements were done from the North. With such unbalanced distribution of users we expect that any systematic deviation in the execution of the experiment, like for example a slight raising of the phone to view the screen, also translates into a North-South deviation.
We have made the raw data with reduced user location accuracy and replaced user ids available for students to work with on our website \cite{suntrace} and it is also available in the supplementary material of this article.

\section*{f. Implementation for other courses}

As demonstrated in these examples, the network interface of phyphox is very versatile and can be adapted for many experiments. Although phyphox is open source (GNU General Public Licence v3) and the app itself could be modified by any user with sufficient skills, there is no need to do so as all examples can be realized within our XML format defining which sensor are read, which mathematical operations to apply to the data and how to transmit it. The file format and the network interface are fully documented on the project’s website.

However, using this interface for a custom experiment in another science course still requires some technical knowledge as the phyphox project cannot offer hosting services for this type of experiment. Therefore, a server to receive the data needs to be set up to at least store incoming data and some basic knowledge of the used protocols (HTTP or MQTT at the time of this article) is necessary. Depending on the target audience, scale and duration of the experiment, this should also include sufficient IT security understanding to not expose the server and the experiment data to attackers. Still, we assess that this type of experiment can be set up in a trusted network environment by anyone with basic web hosting knowledge using PHP. An example is provided in the documentation and supplementary material.

\section*{Conclusion and outlook}

Combining the data from smartphone experiments that are being conducted by multiple learners can open up many exciting and engaging new learning experiences. Integrating a network interface into an app that is used to access the sensors can effectively automate the entire process, especially when paired with automated data analysis and an automated process to present the data. The latter are not strictly necessary, but it allows for an immediate and conclusive feedback and we are convinced that seeing the result developing in real time is an important part of this concept to engage students.

The biggest problem at the moment is a relatively high technical hurdle for the educators as a server and the technical knowledge to set it up are required. The best solution would be offering this concept as a ready-to-use service, but this would require funding for staff and hardware. The next best are examples that should allow teachers in higher education to find an assistant who can use these as templates to set up such an experiment.

Once the technical conditions are met, this concept is very scalable and lifts almost any limits to the number of participants and their location. This allows for collaborative experimentation in remote learning situations, for a community of casual learners around the globe or simply to transform the experience of an entire lecture hall from the purely passive observation of a demonstration experiment to active participation.

\section*{References}

\bibliographystyle{unsrt}
\bibliography{phyphox-network}

\begin{thebibliography}{10}

\bibitem{vogt}
Patrik Vogt, Jochen Kuhn, and Sebastian Müller.
\newblock Experiments using cell phones in physics classroom education: The
  computer-aided g determination.
\newblock {\em The Physics Teacher}, 49(6):383--384, 2011.

\bibitem{pendrill}
Ann-Marie Pendrill and Johan Rohl{\'{e}}n.
\newblock Acceleration and rotation in a pendulum ride, measured using an
  {iPhone} 4.
\newblock {\em Physics Education}, 46(6):676--681, oct 2011.

\bibitem{kuhn}
Jochen Kuhn and Patrik Vogt.
\newblock Smartphones as experimental tools: Different methods to determine the
  gravitational acceleration in classroom physics by using everyday devices.
\newblock {\em European Journal of Physics Education}, 4(1):47--58, 2017.

\bibitem{chevrier}
Joel Chevrier, Laya Madani, Simon Ledenmat, and Ahmad Bsiesy.
\newblock Teaching classical mechanics using smartphones.
\newblock {\em The Physics Teacher}, 51(6):376--377, 2013.

\bibitem{vieyra}
Rebecca Vieyra, Chrystian Vieyra, Philippe Jeanjacquot, Arturo Martí, and
  Martín Monteiro.
\newblock Turn your smartphone into a science laboratory.
\newblock {\em The Science Teacher}, 082, 01 2015.

\bibitem{stampfer}
Christoph Stampfer, Heidrun Heinke, and Sebastian Staacks.
\newblock A lab in the pocket.
\newblock {\em Nature Reviews Materials}, 5(3):169--170, Mar 2020.

\bibitem{Kaps_2021}
A~Kaps, T~Splith, and F~Stallmach.
\newblock Implementation of smartphone-based experimental exercises for physics
  courses at universities.
\newblock {\em Physics Education}, 56(3):035004, feb 2021.

\bibitem{odenwald}
Sten Odenwald.
\newblock Smartphone sensors for citizen science applications: Radioactivity
  and magnetism.
\newblock {\em Citizen Science: Theory and Practice}, 4:18, 05 2019.

\bibitem{lemmens}
Rob Lemmens, Vyron Antoniou, Philipp Hummer, and Chryssy Potsiou.
\newblock {\em Citizen Science in the Digital World of Apps}, pages 461--474.
\newblock Springer International Publishing, Cham, 2021.

\bibitem{staacks}
S~Staacks, S~Hütz, H~Heinke, and C~Stampfer.
\newblock Advanced tools for smartphone-based experiments: phyphox.
\newblock {\em Physics Education}, 53(4):045009, may 2018.

\bibitem{Kaps_2020}
A.~Kaps and F.~Stallmach.
\newblock Using the smartphone as oscillation balance.
\newblock {\em The Physics Teacher}, 58(9):678--679, 2020.

\bibitem{wiki}
{phyphox / RWTH Aachen University}.
\newblock phyphox wiki, 2021.

\bibitem{suntrace}
{phyphox / RWTH Aachen University}.
\newblock Sun trace experiment, 2019.

\end{thebibliography}

\end{document}